\documentclass[showpacs,twocolumn,prl]{revtex4}
\usepackage{amsmath,graphicx}
\usepackage{textcomp,booktabs,tabularx}
\begin{document}

\title{ Full range of proximity effect probed with Superconductor/Graphene/Superconductor junctions }
\author{Chuan Li}\author{S. Gu\'{e}ron}\author{A. Chepelianskii}\author{H. Bouchiat}
\affiliation{Laboratoire de Physique des solides, Univ. Paris Sud }

\begin{abstract}
The high tunability  of the density of states of graphene \cite{Novoselov} makes it an ideal probe of  quantum transport in different regimes. In particular, the supercurrent that can flow through a non-superconducting (N) material connected to two superconducting electrodes, crucially  depends on the lenghth of the N relative  to the superconducting coherence length.
Using graphene as the N material we have investigated  the full range of the superconducting proximity effect, from short to long  diffusive junctions. By combining several S/graphene/S samples with different contacts and lengths, and measuring their gate-dependent critical currents ($ I_c $) and normal state resistance $ R_N $, we compare the product $eR_NI_c$ to the relevant energies, the Thouless energy in long junctions and the superconducting gap of the contacts in short junctions, over three orders of magnitude of Thouless energy. The experimental variations strikingly follow a universal law, close to the predictions of the proximity effect both in the long and short junction regime, as well as in the crossover region, thereby revealing the interplay of the different energy scales.  Differences in the numerical coefficients reveal the crucial role played by the interfacial barrier between graphene and the superconducting electrodes, which reduces the supercurrent in both short and long junctions.
Surprisingly the reduction of supercurrent is independent of the gate voltage and of the nature of the electrodes.   A reduced  induced gap  and Thouless energy are  extracted,  revealing the role played by the dwell time in the barrier in the short junction, and an effective increased diffusion time  in the long junction. We compare our results to the theoretical predictions of Usadel equations and numerical simulations which 
better reproduce  experiments with imperfect NS interfaces.

\end{abstract}

\maketitle

The superconducting proximity effect describes the phenomena that occur when a superconductor (S) is placed in contact with a non-superconducting conductor (\rq\rq{}normal\rq\rq{} conductor, N), and superconducting properties are induced in the N due to the propagation of correlated Andreev pairs from the superconductor to the N \cite{deGennes}. 
Several experiments have revealed the striking effects of induced superconductivity: density of states measurements with tunnel probes have shown how the pair correlations develop as a function of distance to the NS interface \cite{Pothier}; how a minigap is induced in the N when it is connected to two S, and how this minigap is modulated by a magnetic flux that induces a phase difference between the two S electrodes  \cite{Lesueur}. In fact, not only the minigap but the entire energy spectrum of the Andreev eigenstates is phase-dependent, leading to a dissipationless supercurrent that can flow through the normal conductor. The phase dependence of the supercurrent has been probed both at high and low frequency \cite{chi}. 
It is remarkable that all the aforementioned experiments could be described by the theory of the proximity effect, irrespective of the superconducting and normal metals used, their length, aspect ratios, and diffusion constants. The universality of the proximity effects stems from the diffusive motion of carriers, that links the distance pair correlations can travel to a characteristic diffusion time, and hence an energy. 
A striking example of this universality is given by the maximum  supercurrent that can flow through a diffusive SNS junction, called the critical current. The theory of the proximity effect predicts that the critical current is proportional to the smallest correlation energy scale of the problem: the gap energy of the superconducting electrodes if the N is shorter than the superconducting coherence length, or the Thouless energy, proportional to the inverse diffusion time through the N, in the case of long junctions.  The full length dependence of the critical current, from short to long diffusive junctions, was calculated in Dubos \cite{Dubos}, and provides a stringent test of the universality of the proximity effect. This prediction has to our knowledge not yet been tested.

An instance where the universality of the proximity effect could break down is when the interfaces between the N and the S are non ideal. Fermi velocity differences between the N and S materials, disorder,  Shottky,  or insulating barriers at the interface, often cause an interface resistance, that limits the number of induced Andreev pairs, and thus the critical current. The way the critical current is modified in both the short and long limits has been addressed theoretically \cite{Golubov,Brinkman,Cuevas}, but is difficult to test experimentally, since metallic SNS junctions are not tunable. Each junction is unique, with a sample-dependent interface resistance that is often difficult to disentangle from the intrinsic resistance of the normal metal.

In this Letter, we test the universality of the proximity effect in diffusive SNS junctions, from the short- to long-junction regime, over an unprecedented  three order of magnitude range of the Thouless energy over gap ratio. This is  made possible by using graphene as the normal conductor in S/graphene/S (SGS) Josephson junctions. Indeed, graphene\rq{}s carrier density can be controlled by a gate voltage, leading to the possible  continuous spanning, in a single sample of given length  $L$  and aspect ratio, of both the Fermi wave-vector and the diffusion constant $D$, and thus the Thouless energy $E_{Th} =\hbar D/L^2$\cite{metals,previous}.

We track the critical current $I_c$ of seven diffusive graphene samples, using three different superconducting materials, over a wide range of gate voltage.  We find that the product $R_NI_c$  of the critical current  by the  normal state resistance  is proportional to an effective Thouless energy that is a fraction of the Thouless energy,  for all long junctions investigated. In the short junction limit, we find that the $R_NI_c$ product is independent of the Thouless energy, and that 
it is smaller than the electrodes superconducting gap $\Delta$. We find that data of all samples collapse on a single curve, that we compare to the theoretical prediction of the Usadel equations. 
In addition, we perform numerical simulations of the proximity effect in the experimentally relevant situation of an interface with a partial transmission. The simulations  reproduce  qualitatively the  behavior suggested by the experiments, underscoring the role played by multiple inner reflections of Andreev pairs that increase the dwell time in the N conductor.
 
The critical current of short diffusive SNS junctions ($\Delta/E_{Th} \rightarrow 0$) is predicted to obey (\cite{Kulik,Likharev})
\begin{equation}
eR_NI_c \simeq 1.326\pi\Delta/2 \simeq 2.07 \Delta.
	\label{eq_shortjct}
\end{equation}
Whereas the full superconducting gap $\Delta$ is induced in N in short junctions, in long-junctions ($\Delta/E_{Th} \rightarrow \infty$) a much smaller , or  \lq\lq{}mini\rq\rq{} gap  $\Delta_g $  is induced in the N. $\Delta_g$ is proportional to the Thouless energy: $\Delta_g \simeq 3.1 E_{Th}$ \cite{Spivak}.  The product $ eR_NI_c $ at zero temperature is also  proportional to $E_{Th}$ \cite{Dubos}:
\begin{equation}
	eR_NI_c(T=0) = 10.82E_{Th} = 3.2 \Delta_g .
	\label{eq_longjct}
	\end{equation}
 Expressions (1) and (2) show that it is the smallest of the two energies, $\Delta$ and $E_{Th}$,  that limits the critical current in diffusive SNS junctions.  The crossover between the short and  long junction regimes was also investigated using Usadel equations \cite{Dubos}, and it is found that throughout the full proximity effect range, only the diffusive constant, sample length and superconducting gap determine the critical current, regardless of sample geometry. This universality is unique to the diffusive regime: in ballistic SNS junctions, the critical current is  expected to depend on the detailed geometry of the samples \cite{Nazarov}.
Expression (2) was found to reproduce quite well experiments on long metallic SNS junctions \cite{Angers} with a good transmission at the SN interface.  It was however  shown \cite{Brinkman,Golubov,Cuevas} that (1) and (2) are modified by interfacial barriers. The barriers are characterized by  an energy scale $\gamma = \hbar /\tau_\gamma$, with $\tau_\gamma$ the typical time associated with the barrier transmission. The barrier can be  of various types: tunnel, Schottky or due to disorder at the NS interface, a higher barrier  corresponding to a longer dwell time and shorter $\gamma$. When $\gamma$ is smaller than the superconducting gap, $eR_NI_c$  for short junctions is limited by $\gamma$, and independent of $\Delta$ \cite{Brinkman}. The situation is more complex in long junctions with an interfacial barrier such that  $\gamma > E_{Th} $, and was less investigated theoretically.   The interfacial barrier is  often modeled by a simple resistance $R_c \sim h/e^2M\tau$ due to  $M$ conduction channels of identical transmission $\tau < 1$, and characterized by the ratio
$r=R_c/(R_N-2R_c)$, where $R_N$ is the total normal state  resistance, i.e. that of the conductor and barrier resistances in series. In the high r limit, it was found that in short junctions the  induced gap $\Delta^*$ (defined as $eR_NI_c/2.07$) is reduced according to $\Delta^*= \Delta/r$.  In long junctions (\ref{eq_longjct}) is  also  predicted to be modified with a reduction  of the minigap and  critical current   that essentially depends on r and practically   not on   $\tau$ \cite{Cuevas}. 

In the following we present our experimental results on S/graphene/S junctions differing by their superconducting electrodes, length and mobility. Varying the doping changes r substantially,  revealing a  striking universal behavior, and  offering a stringent test of theoretical predictions.

All the samples reported in this paper were prepared by mechanical exfoliation onto oxidized substrates of highly doped Si.  Sample parameters are given in table  \ref{Tab:chapSPE_tableSample}.  The Ti/Al contacts are e-beam evaporated and the Pd/Nb and Pd/ReW contacts are dc-sputtered.   The sample length  L  varies from 300 nm to 1.2$\mu m$ and the ratio $ \xi_s/L $, where $ \xi_s = \sqrt{\hbar D /\Delta } $ is the superconducting coherence length,  varies from 10 to 0.3,  so that the full range from short to long junction is accessed for the first time.

The gate-voltage-dependence of  the normal state resistance $R_N =2R_c +R_G $ yields both the contact resistance $2R_c$  and the  intrinsic graphene resistance $R_G$.   Within a good approximation  $R_G$ is found to vary like $1/|V_g-V_D| $ \cite{Novoselov} at high gate voltage $V_g$ relative to the Dirac point $V_D$. $R_c$  is found to be independent of $V_g$,  and is obtained by the  linear extrapolation of $R_N = f(x=1/|V_g-V_D|) $ close to $x=0$. We can then determine the conductivity  $\sigma =\rho^{-1}= (R _G W/L)^{-1}  =  (2e^2/h) (k_Fl_e)$  and deduce the  mean free path $l_e$, the diffusion coefficient $D=1/2v_Fl_e$, and the Thouless energy. The Fermi wave-vector $k_F$ is deduced from a simple  capacitance model, valid away from the Dirac point \cite{kF}.  The elastic mean free path $l_e$ varies with $V_g$ from 50 nm to 160 nm. Our samples are thus always in  the diffusive   regime. The contact resistance $ R_c$, between  tens and hundreds of Ohms (see table), corresponds to a rather uniform product of contact resistance by sample width, of the order of $250~ \Omega \pm 50 ~\mu m$. $R_c$ is thus  negligible  at low doping, but can be of the order of, or even larger than, the intrinsic resistance of graphene at high doping. This is an ideal parameter range to test the dependence of the proximity effect with$ r$.
Using the expression for the conduction channels $M = k_FW/\pi$, that yields roughly 80 channels for a micron-wide sample at $V_g-V_D=\pm30~V$, one can then deduce the average transmission $\tau$ of the contacts via $R_c= (h/4 e^2)M^{-1}(1/2+(1-\tau)/\tau)$ \cite{Datta}, and we find $\tau= 0.25\pm 0.1$.


\begin{table*}[htp]
	\caption{Characteristic parameters of the  investigated samples. \newline The large contact resistances measured for the Al4 and Al5 samples are  not intrinsic to the sample but due to silver paste connection problems } \centering
	\begin{tabular}{ | p{2cm} | p{.9cm} |p{.9cm} | p{.9cm} |p{.9cm} |p{.9cm} |p{.9cm} | p{.9cm} |p{.9cm} | p{.9cm} |p{.9cm} | p{.9cm} |p{.9cm} | p{.9cm} |p{.9cm}  |}
	\toprule
\hline
	&\multicolumn{10}{c|}{\bf Short/intermediate junction} & \multicolumn{4}{c|}{\bf long junction} \\ \hline
	& \multicolumn{10}{c|}{ Ti/Al(6nm/70nm)}  & \multicolumn{2}{m{1cm}|}{Pd/Nb $8/70~nm$} & \multicolumn{2}{m{.9cm}|}{Pd/ReW $8/70 ~nm$}\\ \hline
	\footnotesize Sample & \multicolumn{2}{c|}{Al1}&\multicolumn{2}{c|}{Al2} & \multicolumn{2}{c|}{ Al3} &\multicolumn{2}{c|}{Al4}&\multicolumn{2}{c|}{\footnotesize Al5}& \multicolumn{2}{c|}{Nb} &\multicolumn{2}{c|}{ReW}\\ \hline
	L (nm) &\multicolumn{2}{c|}{500}&\multicolumn{2}{c|}{400} &\multicolumn{2}{c|}{350}&\multicolumn{2}{c|}{450} & \multicolumn{2}{c|}{500}&\multicolumn{2}{c|}{1200}&\multicolumn{2}{c|}{700} \\
	\hline
	W ($\mu m$)&\multicolumn{2}{c|}{3.4}&\multicolumn{2}{c|}{4}&\multicolumn{2}{c|}{4}&\multicolumn{2}{c|}{4}& \multicolumn{2}{c|}{4}&\multicolumn{2}{c|}{12}&\multicolumn{2}{c|}{5}\\
	\hline
	$\overline{l_e} ~(nm)$& \multicolumn{2}{c|}{120}&\multicolumn{2}{c|}{140}&\multicolumn{2}{c|}{150}&\multicolumn{2}{c|}{120}&\multicolumn{2}{c|}{170}&\multicolumn{2}{c|}{60}&\multicolumn{2}{c|}{70}\\
	\hline
	$\overline{\xi_s} ~ (nm)$&\multicolumn{2}{c|}{400}&\multicolumn{2}{c|}{500}&\multicolumn{2}{c|}{430}&\multicolumn{2}{c|}{420}&\multicolumn{2}{c|}{520}&\multicolumn{2}{c|}{120}&\multicolumn{2}{c|}{120} \\
	
	\hline
	$R_c ~ (\Omega) $&158&136&105&110&172&170&N.A.&860&862&853&40&60&98&120 \\
	\hline
	$V_{g~}(V)$   range&$-35$  $-2.5$&$+16.5$ $+35$&$-35$  $-10$&$+15$       $+25$&$-30$  $-5$&$+5$       $+30$&N.A.&$-10$  $+5$&$-30$    $-20$&$+12$       $+30$&$-30$  $-3$&$+10 $     $+20$&$-25 $  $+ 5$&$+15$     $+ 25$\\
	\hline
	\bottomrule
\end{tabular}
	\label{Tab:chapSPE_tableSample}
\end{table*}

\begin{figure}
	\includegraphics[clip=true,width=9cm]{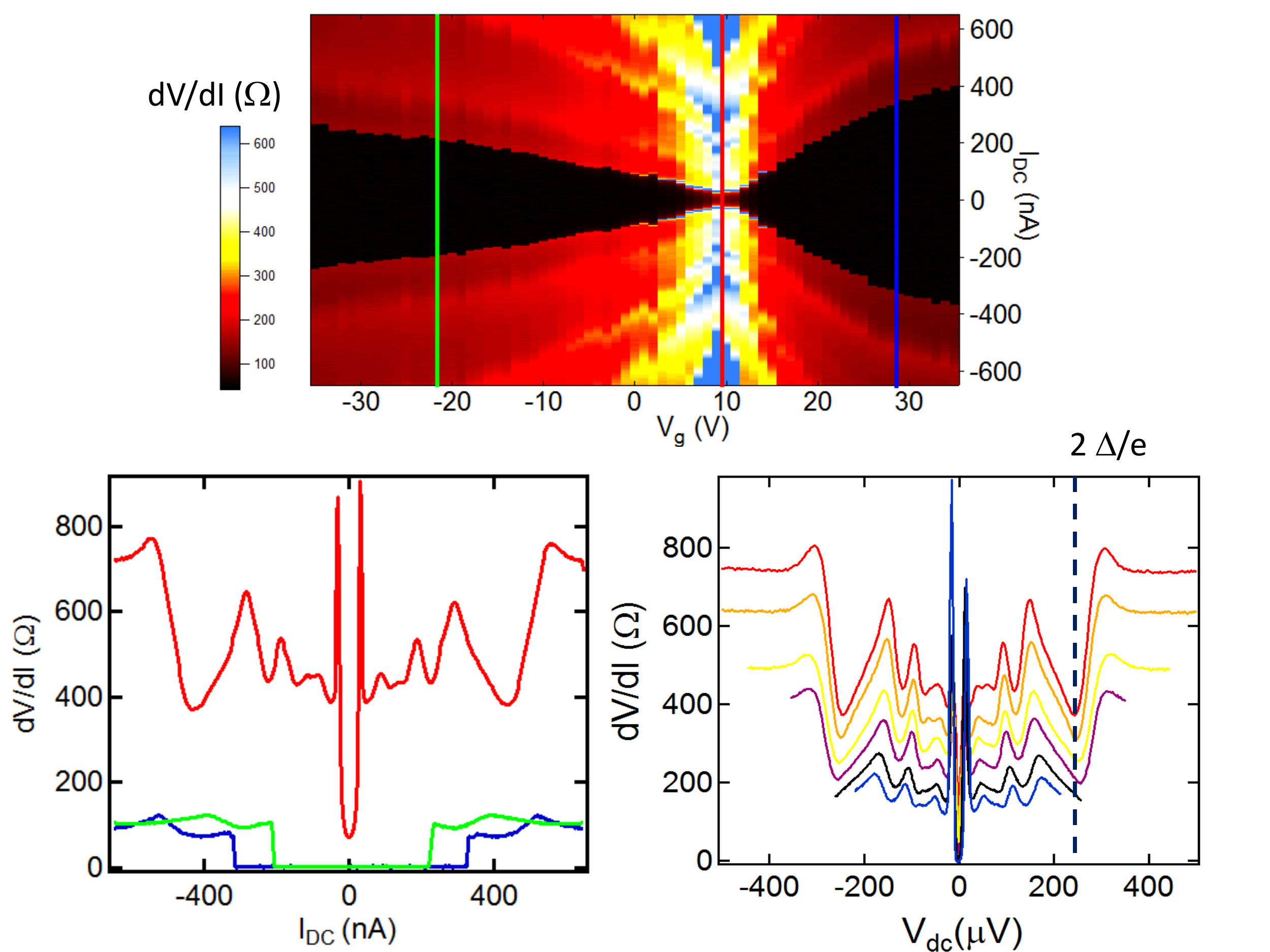}
	\caption{ Top: Color-coded plot of dV/dI  as a function of gate voltage and dc current. Black corresponds to  zero resistance. 
Bottom left: Differential resistance dV/dI($I_{DC}$) at three different gate voltages, including at $V_G$=0V, close to the Dirac point (red curve). The resistance jumps from zero to the normal state resistance at the critical current $I_c$. The peaks in the differential   correspond to Multiple Andreev reflections as clearly seen  on the right plot where the same data is shown as a function of the dc voltage  drop through the sample.
}
	\label{Fig:Fig1_IcVg}
\end{figure}

The differential resistance of the samples was measured at 100 mK via filtered lines, using  a standard lock-in technique.
Fig. \ref{Fig:Fig1_IcVg} displays the colour-coded differential resistance as a function of the bias current and gate voltage, showing a gate-dependent critical current of the SGS junction Al1. The critical current is strongest at high doping, and depressed at gate voltages close to the Dirac point. Peaks in the differential resistance  at $V_n= 2\Delta/ne$ are  manifestations of the multiple Andreev reflections (MAR), typical of SNS junctions \cite{Octavio} and enable the determination of $\Delta$. 

All samples show qualitatively similar behaviors, with quantitative differences: in the long junction samples (Nb, ReW), the critical current is not just depressed, but is actually destroyed near the Dirac point. We attribute this striking suppression to the charge puddles in the sample near half filling, and the specular Andreev reflections across their boundaries, that randomize the phase of Andreev pairs \cite{Komatsu2012}. In the following we focus on data sufficiently far from the Dirac point  (Table I) so that the critical current is higher than 100 nA. This ensures that  thermal fluctuations have a negligeable influence, since the corresponding Josephson energy $E_J = \Phi_0 I_c/2\pi$ is above 3 K, more than ten times the sample temperature\cite{Ambegaokar}. We show in the following that all samples exhibit  a universal behavior.

\begin{figure}
    \includegraphics[clip=true,width=9cm]{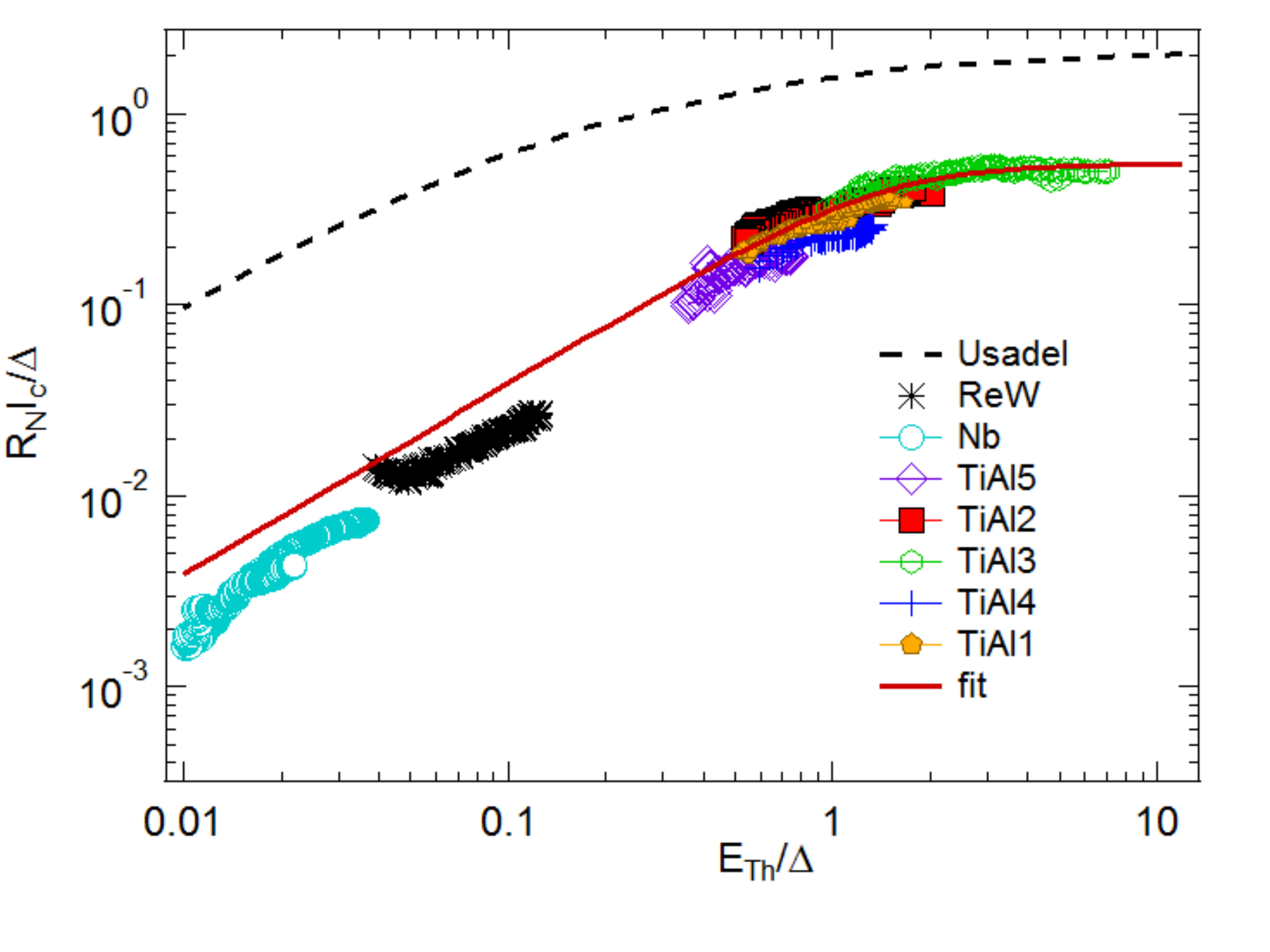}
	\caption{Variations of  $ R_NI_c/\Delta$  with  $x= E_ {Th}/\Delta$ for seven diffusive SGS junctions, of different lengths and with different superconducting electrodes. The superconductor used as a contact is indicated in the legend. For each sample, a continuous range of Thouless energy is accessed by varying the gate voltage. Three orders of magnitude of Thouless energy over Delta are accessed. All data  practically collapse on a single curve $F(x)$ (red continuous line), see expression (3), whose shape describes both the short and long junction limits, as well as the crossover between the two regimes. This shape is similar to the theoretical curve for a perfect interface, computed from the Usadel equations by Dubos et al.\cite{Dubos}  (dashed  black curve).}
	
    \label{Fig:normalization} 
	\end{figure}

To follow and compare the critical current of all samples, we plot the experimentally determined $eR_NI_c/\Delta $ as a function of $x=E_{Th}/\Delta$,   
(Fig. \ref{Fig:normalization}), along with the numerical solution $F_U(x)$ of the Usadel equations for perfect interfaces \cite{Dubos}.
We find that all experimental data nearly collapses on a single curve, $eR_NI_c/ \Delta=F(x)$,  with two  asymptotic behaviors  that clearly correspond to the long and short junction limits. This universal behavior of all the  graphene-based SNS junctions we have investigated is the central result of our paper.  In the short junction limit,  $ \lim _{x\rightarrow  \infty }F(x) = a $	  with $a = 0.55 $, and  in the long junction limit $\lim _{x \rightarrow  0}  F(x) =  bx $ with $b\simeq 0.39 $. This behavior is qualitatively similar to the result of Usadel equations, although the Usadel equation coefficients are different:  $a_U \simeq 2.07  $  and  $ b_U= 10.82 $ \cite{Dubos}.
This comparison with Usadel equations leads us to define effective energies $\Delta^* = (a/a_U) \Delta \simeq 0.3 \Delta $ and  $E_{Th}^*  = (\Delta/\Delta*)(b/b_U) E_{Th} \simeq 0.14 E_{Th} $ such that $ eR_NI_c/ \Delta^*\rightarrow  a_U$ in the short junction limit and 
$ eR_NI_c/ \Delta^*\rightarrow b_U E_{Th}^*/\Delta$ in the long junction limit. The full dependence, including  the crossover between short and long junctions, can be fitted by a generic expression:	 

\begin{equation}
	\frac{eR_NI_c}{\Delta}=F(x)=  \frac{ abx } {(a^n + b^n x^n)^{1/n}}.
	\label{crossover}
	\end{equation}
Fig. 2 shows that the  Usadel   results are very well fitted by $n \simeq 1$, and  the experiments, with a sharper crossover, by $n \simeq 2$.  Fig. 2 also shows that an imperfect interface does not change the main features of the proximity effect: in short junctions $R_N I_c$ is independent of $E_{Th}$, and in long junctions $R_N I_c$ varies linearly with $E_{Th}$.
	   According to refs.\cite {Brinkman,Golubov},  the reduced effective gap $\Delta^*$ should just be $\gamma$, the inverse characteristic  transmission time through the NS barrier in the limit where $\Delta \gg \gamma \gg E_{Th}$. 
It is interesting that the $\Delta^*$ we find is sample independent, for the three samples with Ti/Al contacts for which the crossover between long and short junction is accessed. One would have liked to test samples with different superconducting gaps in this short junction regime to determine whether $ \Delta ^* = \gamma$  is independent of $\Delta$. This was not possible for the Pd/ReW and Pd/Nb contacts, whose very small superconducting coherence length would require sub 30-nm-size junctions to reach the short junction limit. 
 
 The physical meaning of the crossover between short and long junctions in the presence of barriers  can be heuristically understood writing that  $eR_N I_C = \hbar/ \tau_{dw}$, where $\tau_{dw}$ stands for the typical traversal time of the SNS junction, which is the sum of $\tau_\gamma$, the time spent in the barriers, and $\tau_D$, the  diffusion time through the normal junction, yielding $eR_N I_c= \gamma E_{Th}/(E_{Th} +\gamma)$. This expression reproduces quite well the solution of the Usadel equations \cite{Dubos}, with $\gamma$ instead of $\Delta$, and corresponds to eq. (\ref{crossover}) with n=1. Its dependence is similar to the experimental curve, although the theoretical crossover is smoother than the experimental curve, that is better described by n=2.

We now turn to the long junction regime, in which  the critical current varies linearly with $E_{Th}$, but is smaller than the theoretical prediction for perfect interface by a factor $b_U/b\simeq 33 =  3 E_{Th}/E_{Th}^*$ \cite{Dubos}.	 
In \cite{supp} we show that this reduced Thouless energy $E_{Th}^*$ also determines the temperature dependence of $I_c$.

Modifications of  $I_c(0)$ and  $I_c(T)$ due to imperfect interfaces were investigated  by  Hammer et al. \cite{Cuevas} using the Usadel equations formalism. They predict  that the  renormalised critical current and its variations with temperature depend not only on $E_{Th}/\Delta $ but also on  $r= R_c/(R_N -R_c)$,  with a drastic reduction of $R_NI_c$ at high $r$. Since $r$ varies in graphene by a factor $50$ ($r\simeq0.1$ close to $V_g=V_D$, and $r\simeq5$ around $V_g\simeq30 ~ V$), one would expect $E_{Th}^*/E_{Th}$ to vary with doping, in strong contrast to the  universal behavior suggested from our data.
The same calculation also predicts a critical current that is  only barely reduced so long as the interface resistance is small relative to the normal conductor\rq{}s resistance ($r\ll1$). 
For $r= R_c/R_N \simeq 0.1$ for instance, as in our experiments at low doping, the prediction would be $b/b_U \simeq 1$ (using our notations). This is in stark contrast with our experimental finding of $b/b_U  \simeq 0.03$.  Such a high reduction due to a relatively small interface resistance was already reported by Dubos et al.\cite{Dubos} in a metal SNS junction. 

A possible interpretation of a strongly reduced effective Thouless energy ($E_{Th}^* \ll E_{Th}$) could be the repeated inner reflections  of Andreev pairs at the interfacial barriers, leading to an increased typical time spent in the SNS junction from $\tau_D$ to $  N\tau_D$, where N is the number of reflections at the NS interfaces.

We now present numerical simulations, in which we do find a strongly (ten-fold) reduced critical current, even for $r\ll1$, i.e. a graphene sheet whose  intrinsic resistance is  much higher than the interface resistance. We implement the Bogoliubov-de Gennes Hamiltonian 
that describes the electron- and hole-like wavefunction components of a hybrid  NS ring in a tight-binding 2D Anderson model \cite{Ferrier}. 
The graphene sheet  is a hexagonal lattice oriented along the armchair direction with  $N_x\times N_y$  sites, and is connected to two superconducting electrodes  ($N^S =N_x^S \times N_y $ sites on a square lattice), see inset of Fig. 3. Disorder is described by random on-site energies  of variance $W^2$. The hopping matrix element  is restricted to nearest neighbors   $t_{ij}=t $. The SN interface barrier is taken into account via a reduced hopping amplitude between the N and S sites: $|t_{SN}/t|^2 =\tau $, with $0<\tau<1$, \cite{parameters}
 The Josephson current $I_J (\varphi) = \partial E_J/\partial\varphi$ is  the derivative of the Josephson energy $E_J$, the sum of the  occupied, phase-dependent energy levels. 
The critical current $I_c$ is the maximum of $I_J (\varphi)$. Fig. 3  displays the  length dependence of $I_c$ of a graphene ribbon with $N_y=60$   and the   dependence of $R_NI_c$  with Thouless energy. When $\tau=1$ (perfectly transmitting interfaces), we find that $I_c$ varies as  $1/L$ in short junctions, i.e. for $L$ smaller than the superconducting coherence length (of the order of 10 lattice spacings), in accordance with Eq. (1), $eR_NI_c \simeq 2.07 \Delta$. For long junctions, a faster $1/L^3$ decay is observed, in accordance with Eq. (2), $eR_NI_c \simeq 10.8E_{Th}$.  
$R_c$ and $R_N$ are estimated via  $ R_c= (1-\tau)/N_y\tau +1/N_y$ and $R_N = 2R_c + L /N_y l_e$. As expected, the simulation displays an approximately linear relation between $R_NI_c$ and $E_{Th}$ in the long junction regime. 
 More crucially, the simulations also show a striking reduction in the critical current of long junctions for an interface transmission $\tau$ of 0.12, even when r is smaller than 0.1. We also find this effect for a square lattice instead of a hexagonal lattice, demonstrating that it is  not specific to graphene. This strong reduction of supercurrent  by a relatively small resistance barrier, that we find in the experiment and in the simulations, is a central result of our paper. Such a reduction of the supercurrent  is to our knowledge  not predicted in \cite{Cuevas}. 
\begin{figure}
	\includegraphics[clip=true,width=9cm]{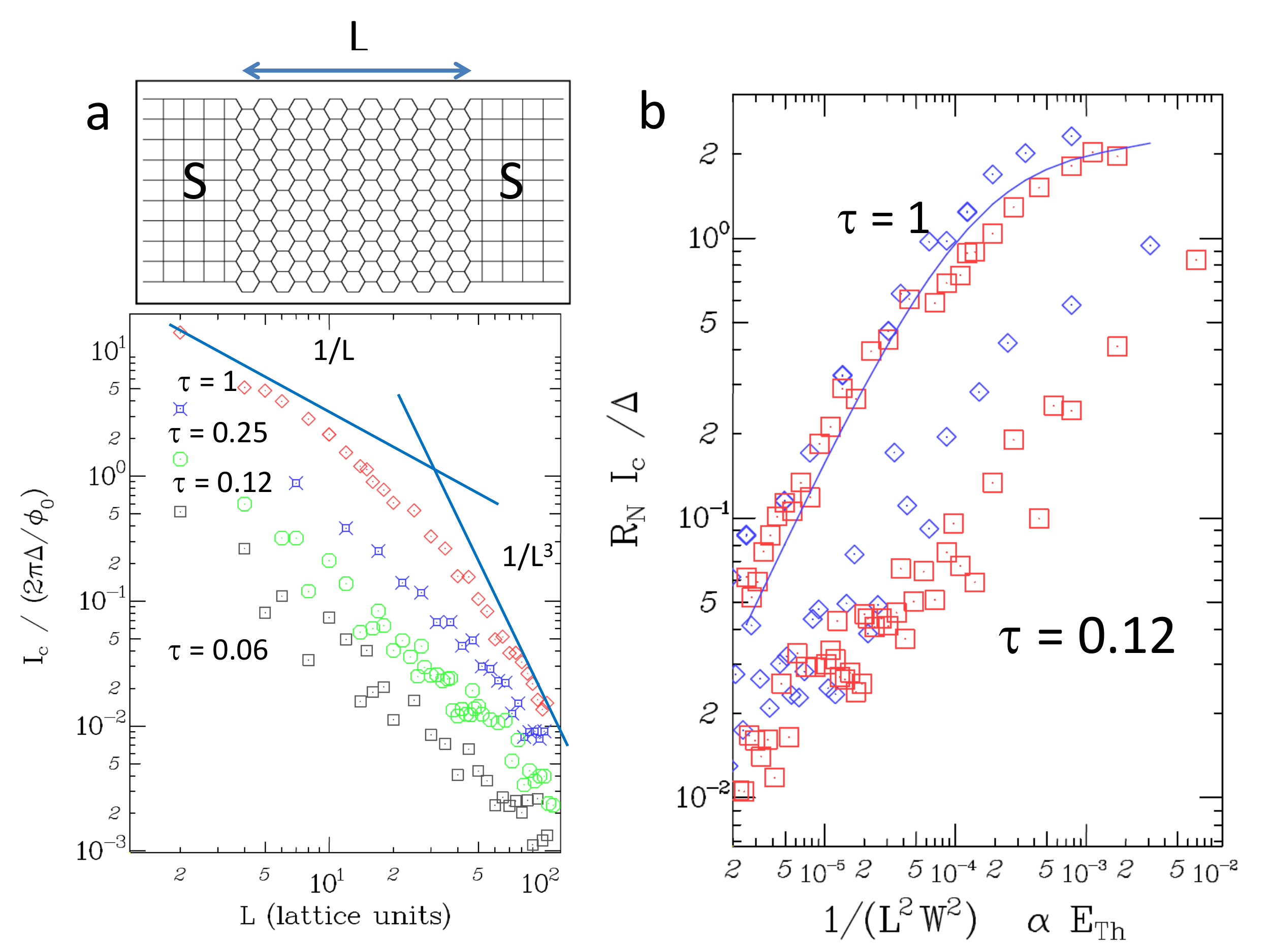}
	\caption{ a- Length dependence of the Josephson current calculated for a diffusive  graphene ribbon between superconducting electrodes for different values of $\tau$. The number of transverse channels is $N_y=60$, the disorder is  $W=6$.  
	b-The product  $R_N I_c$ is shown as a function of $1/ W^2N_x^2$  (that is proportional to the Thouless energy), for 2 different values of disorder (W=6, red symbols,   and W=9, blue symbols.) }
	\label{Fig3}
\end{figure}

In conclusion, we have tested the Thouless energy dependence of the critical current of diffusive SNS junctions over three orders of magnitude, thanks to the tunability of graphene used as the diffusive normal conductor. Our analysis of the critical current in different graphene-based Josephson junctions in the diffusive regime  shows  a remarkable universal behavior, with a crossover between long and short junctions regimes. The full dependence of $R_NI_c$ Vs $E_{Th}/\Delta$ can be described by the result of the Usadel theory with perfect interfaces, provided  a sample-independent rescaling of the superconducting gap and Thouless energy down to lower energies is performed. We understand this reduction  as due to the  barriers at the NS interface, whose transmission is estimated to be of the order of 0.25. We find that the  predictions of Usadel equations in the long junction limit with opaque  interfaces do not agree with the universal behavior we observe. A better agreement is obtained with numerical computation of the Andreev spectrum  using a tight binding model of graphene.  These  results call for a better  theoretical understanding of the influence of  barriers at the N/S interface on the transmission of Andreev pairs through long SNS junctions.

We acknowledge discussions with Richard Deblock, Anil Murani  and Juan Carlos Cuevas as well as CNRS, ANR Supergraph and DIRACFORMAG   for funding.


\begin{thebibliography}{99} 

\bibitem{Novoselov} A. H. Castro Neto, F. Guinea, N. M. Peres, K. S. Novoselov, and A. K. Geim, {\it The electronic properties of graphene}, Rev. Mod. Phys. {\bf 81}, 109 (2009).

\bibitem{deGennes} P.G. de Gennes,{\it Boundary Effects in Superconductors}, Rev. Mod.Phys. {\bf 36},225 (1964). Kulik, I.,{\it Macroscopic quantization and proximity effect in S-N-S junctions},

Sov. Phys. JETP {\bf 30}, 944 (1970).

\bibitem{Pothier} S. Gu\'eron, H. Pothier, Norman O. Birge, D. Esteve, M.H. Devoret,{\it Superconducting proximity effect probed on a mesoscopic length scale}, Phys. Rev. Lett. {\bf 77}, 3025 (1996).

\bibitem{Lesueur}H. le Sueur et al.,{\it Phase Controlled Superconducting Proximity Effect Probed by Tunneling Spectroscopy}, Phys. Rev. Lett. {\bf100}, 197002 (2008).

\bibitem{chi}M. Fuechsle et al.,{\it Effect of Microwaves on the Current-Phase Relation of Superconductor�Normal-Metal�Superconductor Josephson Junctions} Phys. Rev. Lett. {\bf102}, 127001 (2009).

B. Dassonneville, M.Ferrier, S. Gu\'eron, and H. Bouchiat,{\it Dissipation and Supercurrent Fluctuations in a Diffusive Normal-Metal�Superconductor Ring}, Phys. Rev. Lett. {\bf110} 217001 (2013).

\bibitem{Kulik} I. O. Kulik and A. N. Omel\rq{}yanchuk,{\it Contribution to the microscopic theory of the Josephson effect in superconducting bridges}, Zh. Eksp. Teor. Fiz. Pis. Red. 21, 216 [JETP Lett. 21, 96 (1975)].

\bibitem{Likharev} Likharev, K. K.,{\it Superconducting weak links}, Rev. Mod. Phys. 51, 101 (1979).

\bibitem{Golubov} A. A. Golubov, M. Yu. Kupriyanov and E. Illichev,{\it The current-phase relation in Josephson junctions}, Rev. Mod. Phys. 76, 418 (2004).

\bibitem{Dubos} P. Dubos, H. Courtois, B. Pannetier, F. K. Wilhelm, A. D. Zaikin, and G. Sch\"on,{\it Josephson critical current in a long mesoscopic SNS junction}; Phys. Rev. B {\bf 63}, 064502 (2001); P. Dubos, H. Courtois, O. Buisson, and B. Pannetier,{\it Coherent Low-Energy Charge Transport in a Diffusive S-N-S Junction}, Phys. Rev. Lett. {\bf 87}, 206801 (2001). P. Dubos PhD Thesis (2001).

\bibitem{Brinkman} Brinkman, A., and A. A. Golubov,{\it Coherence effects in double-barrier Josephson junctions}, Phys. Rev. B 61, 11297 (2000).

\bibitem{metals} The length dependence of the critical current was investigated in semiconducting nanowires in \cite{Abay}, with mainly short junctions, and metallic wires in \cite { Palevski}, in the limit $r \gg 1$.

\bibitem{Abay} S. Abay, D. Persson, H. Nilsson, F. Wu, HQ Xu, M. Fogelstrom, V. Shumeiko,{\it Charge transport in InAs nanowire Josephson junctions}, Phys. Rev. B 89, 214508 (2014).

\bibitem{Palevski} Y. Blum, A. Tsukernik, M. Karpovski, and A. Palevski ,{\it  Critical current in Nb-Cu-Nb junctions with nonideal interfaces}, Phys. Rev. B 70, 214501 (2004).

\bibitem{previous} Previous work on graphene based SNS junctions mostly focused on the short junction regime \cite{Heersche,Andrei,Girit, Ojeda,Finkelstein,Miao,Coskun,Mizuno}. The long junction diffusive regime was less but also investigated with graphene connected to Pb and Nb electrodes \cite{ Doh, Komatsu2012} and very recently\cite{Calado,Shalom} long ballistic junctions with MoW and Nb electrodes with very high transparency were investigated.

\bibitem{Heersche} H. B. Heersche, P. Jarillo-Herrero, J. B. Oostinga, L. M. K. Vandersypen, and A. F. Morpurgo,{\it Bipolar supercurrent in graphene}, Nature 446, 56 (2007).

\bibitem{Andrei} X. Du, I. Skachko, and E. Y. Andrei,{\it Josephson current and multiple Andreev reflections in graphene SNS junctions}, Phys. Rev. B 77, 184507 (2008).

\bibitem{Girit}C. Girit et al.{\it Tunable graphene dc superconducting quantum interference device}, Nano Lett. 9, 1980 (2009).

\bibitem{Ojeda} C. Ojeda-Aristizabal, M. Ferrier, S. Gu\'eron, and H. Bouchiat,{\it Tuning the proximity effect in a superconductor-graphene-superconductor junction}, Phys. Rev. B 79, 165436 (2009).

\bibitem{Finkelstein} I. V. Borzenets, U. C. Coskun, S. J. Jones, and G. Finkel-
stein,{\it Phase Diffusion in Graphene-Based Josephson Junctions} Phys. Rev. Lett. 107, 137005 (2011).
\bibitem{Miao} F. Miao, W. Bao, H. Zhang and C.N. Lau, {\it Premature switching in graphene Josephson transistors},Solid State Commun.149 , 1046 (2009).
\bibitem{Coskun} U. C. Coskun, M. Brenner, T. Hymel, V. Vakaryuk, A. Levchenko, A. Bezryadin. {\it Distribution of supercurrent switching in graphene under proximity effect.} Phys. Rev. Lett. 108, 097003 (2012).
\bibitem{Mizuno}N. Mizuno, B. Nielsen, X. Du. {\it Ballistic-like supercurrent in suspended graphene Josephson weak links.} Nature Commun. 4, 3716 (2013).
\bibitem{Doh}  G.H. Lee, D. Jeong, Jae-Hyun Choi, Yong-Joo Doh, and Hu-Jong Lee, {\it Electrically Tunable Macroscopic Quantum Tunneling in a Graphene-Based Josephson Junction}, Phys. Rev. Lett. 107, 146605 (2011); Dongchan Jeong, Jae-Hyun Choi, Gil-Ho Lee, Sanghyun Jo, Yong-Joo Doh, and Hu-Jong Lee, {\it Observation of Supercurrent in PbIn-Graphene-PbIn Josephson Junction} Phys. Rev. B 83, 094503 (2011).
\bibitem{Komatsu2012}Katsuyoshi Komatsu, Chuan Li, S. Autier-Laurent, H. Bouchiat, and S. Gu\'eron {\it Superconducting proximity effect in long superconductor/graphene/superconductor junctions: From specular Andreev reflection at zero field to the quantum Hall regime} Phys. Rev. B 88, 115412 (2012).

\bibitem {Calado} Victor E. Calado, Srijit Goswami, Gaurav Nanda, Mathias Diez, Anton R. Akhmerov, Kenji Watanabe, Takashi Taniguchi, Teun M. Klapwijk, Lieven M. K. Vandersypen {\it Ballistic Josephson junctions in edge-contacted graphene.}, Nat. Nano. 10, 761 (2015).
\bibitem{Shalom}
M. Ben Shalom, M. J. Zhu, V. I. Fal'ko, A. Mishchenko, A. V. Kretinin, K. S. Novoselov, C. R. Woods, K. Watanabe, T. Taniguchi, A. K. Geim, J. R. Prance {\it Proximity superconductivity in ballistic graphene, from Fabry-Perot oscillations to random Andreev states in magnetic field}, Nature Physics (2015) (advanced online publication).
\bibitem{kF}The carrier density  induced by the gate voltage is given by:
$n_c = \dfrac{\varepsilon_r\varepsilon_0 }{ed}\lvert V_G - V_D\rvert$,
where   $\varepsilon_r = 3.7 $ for $SiO_{2}$, $ \varepsilon_0 = 8.85 \times 10^{-12} F \cdot m^{-1} $, and  $d = 285 ~nm$ is the thickness of the oxide layer. $k_F$  is  deduced from $n_c$  via $k_F = \sqrt{\pi n_C} \simeq 4.75 \times 10^7 \times \sqrt{V_G - V_D}  ~m^{-1}$.
\bibitem{Spivak} F Zhou, P Charlat, B Spivak, and B Pannetier. {\it Minigap in a
long disordered SNS junction: Analytical results.} Physical Review B,
{\bf 66}, 052507, (2002).
\bibitem{Nazarov}A. Lodder,  Yu. V. Nazarov {\it Density of states and the energy gap in Andreev billiards} PRB 58 5783 (1998).
\bibitem{Angers}L. Angers, F. Chiodi, G. Montambaux, M. Ferrier, S. Gu\'eron, H. Bouchiat,
and J. Cuevas. {\it Proximity dc squids in the long-junction limit}. Physical Review B, 77, 165408, (2008).

\bibitem{Cuevas}J. C. Hammer, J. C. Cuevas, F. S. Bergeret, W. Belzig,{\it Density of states and supercurrent in diffusive SNS junctions: Roles of nonideal interfaces and spin-flip scattering} Phys. Rev. B 76, 064514 (2007).
\bibitem{supp} Supplemental material.
\bibitem{Octavio} K. Flensberg, J. Bindslev Hansen, and M. Octavio, {\it Subharmonic energy-gap structure in superconducting weak links} Phys. Rev. B, 88, 8707 (1988).
\bibitem{Ambegaokar} V. Ambegaokar and B.I. Halperin  {\it Voltage Due to Thermal Noise in the dc Josephson Effect.}Phys. Rev. Lett. {\bf 22}, 1364,(1969).

\bibitem{Datta} Supryo Datta ''Electronic Transport in Mesoscopic Systems'', Cambridge University Press (1995).
\bibitem{Wilhelm} F.K. Wilhelm, A.D. Zaikin, and G. Sch\"on, {\it Supercurrent in a mesoscopic proximity wire}J. Low Temp. Phys. 106, 305 (1997).
\bibitem{Ferrier}M. Ferrier, B. Dassonneville, S. Gu\'eron, and H. Bouchiat {\it Phase-dependent Andreev spectrum in a diffusive SNS junction: Static and dynamic current response}
Phys. Rev. B 88, 174505 (2013).
\bibitem {parameters}  We have chosen the superconducting gap $\Delta =t/4$ such that the S coherence length $\xi_s= a t/\Delta \ll N_x^S$, to avoid suppressed superconducting correlations in the S (inverse proximity effect). The Fermi energy was chosen  away from the Dirac point, at filling 1/4 . The disorder amplitude and $l_e$ are related via $l_e = \alpha a (t/W)^2$ in  2D  \cite{Sigetti}. The number of  transverse channels  and disorder amplitude correspond to  the diffusive regime, for which the normal region length $N_x a$  is greater than the elastic mean free path $l_e$ and shorter than the localization length $N_y l_e$.  T The coefficient $\alpha$ relating $l_e$ to the amplitude of the disorder is determined from the  1/L dependence of the critical current  in the short junction limit at $\tau=1$.

\bibitem{Sigetti}G. Montambaux, H. Bouchiat, D. Sigeti, and R. Friesner, {\it Persistent currents in mesoscopic metallic rings: Ensemble average}, Phys. Rev.B 42, 7647 (1990).
\end{thebibliography}
\end{document}